\documentclass{aa}  
\usepackage[english]{babel}
\usepackage[utf8]{inputenc}
\usepackage[T1]{fontenc}
\usepackage{graphicx}
\usepackage{amsmath,amssymb}
\usepackage{txfonts} 
\usepackage{etoolbox}
\usepackage{wasysym}
\usepackage{threeparttable}
\usepackage{natbib}

\defcitealias{Ofek2014ApJ...789..104O}{O14}
\defcitealias{Bilinski2015MNRAS.450..246B}{B15}
\defcitealias{Strot2021ApJ...907...99S}{S21}
\newcommand{\msun}{M$_{\odot}$ }
\newcommand{\lsun}{L$_{\odot}$}
\newcommand{\kms}{km s$^{-1}$}

\begin{document} 

\title{Searching for precursor activity of Type IIn Supernovae}

\titlerunning{Precursors of SNe IIn}

\author{A. Reguitti\inst{1,2,3,4}\fnmsep\thanks{E-mail: andreareguitti@gmail.com},
G. Pignata$^{5}$, A. Pastorello$^{2}$, R. Dastidar$^{3,4}$, D. E. Reichart$^{6}$, J. B. Haislip$^{6}$, V. V. Kouprianov$^{6}$
}

\authorrunning{Reguitti et al.} 

\institute{
INAF – Osservatorio Astronomico di Brera, Via E. Bianchi 46, 23807, Merate (LC), Italy
\and
INAF – Osservatorio Astronomico di Padova, Vicolo dell'Osservatorio 5, 35122 Padova, Italy
\and 
Instituto de Astrof\'{i}sica, Universidad Andres Bello, Av.da Rep\'{u}blica 252, 8320000 Santiago, Chile 
\and
Millennium Institute of Astrophysics, Nuncio Monsenor S\'{o}tero Sanz 100, Providencia, 8320000 Santiago, Chile
\and
Instituto de Alta Investigaci\'{o}n, Universidad de Tarapac\'{a}, Casilla 7D, Arica, Chile
\and
University of North Carolina at Chapel Hill, Campus Box 3255, Chapel Hill, NC 27599-3255, USA
}

\date{Received XXX; accepted YYY}
 
\abstract
{We conducted a search for luminous outbursts prior to the explosion of Type IIn Supernovae (SNe~IIn). We built a sample of 27 objects spectroscopically classified as SNe IIn, all located at $z<0.015$. Using deep archival SN fields images taken up to nearly 20 years prior from transient surveys (PTF, ZTF, DES, CHASE) and major astronomical observatories (ESO and NOAO), we found at least one outburst years to months before the explosion of seven SNe IIn, the earliest precursor being 10 years prior to the explosion of SN 2019bxq.
The maximum absolute magnitudes of the outbursts range between -11.5 mag and -15 mag, and the eruptive phases last for a few weeks to a few years. 
The $g-r$ colour measured for three objects during their outburst is relatively red, with $g-r$ ranging between 0.5 and 1.0~mag. This is similar to the colour expected during the eruptions of Luminous Blue Variables.
We noticed that the SNe with pre-SN outbursts have light curves with faster decline rates than those that do not show pre-SN outbursts.
SN 2011fh is remarkable, as it is still visible 12 years after the luminous SN-like event, indicating that the progenitor possibly survived, or that the interaction is still on-going.
We detect precursor activity in 29\% of bona-fide SNe~IIn in our sample.
However, a quantitative assessment of the observational biases affecting the sample suggests that this fraction underestimates the intrinsic precursor occurrence rate.
}


\maketitle
%

\section{Introduction}

Supernovae (SNe) IIn are a class of explosive astrophysical events whose spectra show evidence of interaction between the fast SN ejecta and the slow, high-density surrounding material. The interaction is marked by the presence of narrow emission lines of the Balmer series, whose profiles are composed by a narrow component, sometimes with a P Cygni profile, on top of a broader one \citep{Schlegel1990MNRAS.244..269S, Filippenko1997ARA&A..35..309F, Fraser2020RSOS....700467F}. The narrow component arises from a slow-moving ($\sim 10^2$ \kms) circumstellar medium (CSM) produced by the progenitor during previous mass-loss events, and piled up around it. The broad component is generated by the fast-moving ($\sim 10^4$ \kms) SN ejecta. The collision of the ejecta with the CSM generates two shock fronts, one moving inwards through the ejecta and the other outwards through the CSM, with the shocked material producing high energy photons that ionise the CSM. The narrow spectral features are produced from the recombination of the slow-moving ionised gas.
Other typical characteristics of SNe IIn are their high luminosity, produced by the efficient conversion of the ejecta kinetic energy into radiation, and the blue colours, sometimes accompanied by a UV/X-ray excess, due to high temperature of the gas \citep{Chevalier1994ApJ...420..268C}.

In numerous cases the progenitors of SNe IIn showed signatures of strong variability, including recurrent outbursts, in the years before the explosion \citep[for instance][]{Smith2010AJ....139.1451S, Ofek2013Natur.494...65O, Tartaglia2016MNRAS.459.1039T, EliasRosa2016MNRAS.463.3894E, Thone2017A&A...599A.129T, Pastorello2018MNRAS.474..197P, Elias-Rosa2018MNRAS.475.2614E, Pastorello2019A&A...628A..93P, Reguitti2019MNRAS.482.2750R}. The eruptive phases last from a few weeks to a few years \citep[see][]{Smith2011MNRAS.415..773S}. 
These pre-SN events belong to the stellar transient category of ‘SN Impostors' \citep{VanDyk2000PASP..112.1532V}, since their spectra and evolution in some cases can resemble those of a Type IIn SN. However, they are less luminous, with absolute magnitudes ranging between $-10$ and $-15$ mag, and belong to the population of the so called ‘Gap Transients' \citep{Kasliwal2012PASA...29..482K,Pastorello2019NatAs...3..676P,universe8100493}.
Eruptive mass-loss events from Luminous Blue Variables \citep[LBV,][]{Humphreys1994PASP..106.1025H} are a plausible explanation of the SN Impostors phenomenon (\citealt{Gal2007ApJ...656..372G, Gal2009Natur.458..865G}, but see \citealt{Smith2017hsn..book..403S}). 

\cite{Ofek2014ApJ...789..104O} (hereafter \citetalias{Ofek2014ApJ...789..104O}) conducted a systematic search for precursor transient events that later produced Type IIn SNe. They inspected archival Palomar Transient Factory \citep[PTF;][]{Law2009PASP..121.1395L} images for 16 objects, taken between 1 and 3 years before the explosion, and found luminous outbursts for six of them. From this result, they also estimated that about 50\% of Type IIn SNe are expected to experience at least one luminous outburst (brighter than $M_V=-14$ mag) within 4 months before the terminal explosion.
In contrast, \cite{Bilinski2015MNRAS.450..246B} (from now on \citetalias{Bilinski2015MNRAS.450..246B}) conducted a similar search for 6 SNe IIn on archival Lick Observatory Supernova Search \citep[LOSS;][]{Filippenko2001ASPC..246..121F} data from the 76-cm Katzman Automatic Imaging Telescope (KAIT) telescope taken up to 12 years prior, but they found no precursors. This latter result can be explained by the limited sample and the shallow limit magnitude the survey can reach (around 19 mag).

More recently, \cite{Strot2021ApJ...907...99S} (\citetalias{Strot2021ApJ...907...99S}) performed a similar search on a much larger sample of objects (196 interacting SNe) over images from the Zwicky Transient Factory \citep[ZTF;][]{Masci2019PASP..131a8003M} Survey taken between March 2018 and June 2020. They retrieved precursor eruptions prior to 18 Type IIn SNe and one Type Ibn SN. They found that the precursor events become brighter and more frequent in the last months before the SN, and a quarter of all Type IIn SNe experience month-long outbursts brighter than an absolute magnitude of $-13$ mag within the final three months before the explosion. The radiative energies of the outbursts are up to $10^{49}$ erg.

In this paper, we present the results of our independent search for pre-SN outbursts conducted on a sample of 27 Type IIn SNe discovered within about 60 Mpc. We made use of archival images from major astronomical observatories (ESO and NOAO) as well as frames collected by public photometric surveys, such as PTF, ZTF and DES. Our research differentiates from previous ones for the contemporary use of archival data from both surveys and observatories, the latter of which can be much more profound (down to $\sim$22.5 mag), allowing to detect outbursts as faint as $M_r\sim-11$ mag, and for the longer time window monitored (up to 10 years before the SN explosion).

The structure of the paper is the following: in Sect. \ref{sec:sample}, we present the sample of Type IIn SNe and the adopted selection criteria. In Sect. \ref{sec:method}, we describe the methodology of the investigation and the adopted data reduction techniques. The principal results are outlined in Sect. \ref{sect:results}, followed by a discussion in Sect. \ref{sec:discus}. Finally, our conclusions are presented in Sect. \ref{sec:conclus}.


\section{The Sample} \label{sec:sample}
We searched for all objects classified as a Type IIn SN in the Transient Name Server (TNS\footnote{https://www.wis-tns.org/}), Weizmann Interactive Supernova Data Repository (WISeREP\footnote{https://wiserep.weizmann.ac.il/}, \citealt{Yaron2012PASP..124..668Y}) and Asiago Supernova Catalogue\footnote{http://graspa.oapd.inaf.it/asnc.html} \citep{Barbon1989A&AS...81..421B,Barbon1999A&AS..139..531B} databases, up until April 2021, and with a redshift $z<0.015$, retrieving 82 SNe. Assuming a standard cosmology ($H_0=73$ \kms Mpc$^{-1}$, $\Omega_m=0.27$, $\Omega_{\Lambda}=0.73$; these cosmology parameters will be adopted throughout this paper), 
this corresponds to a distance of about 61.6 Mpc, or a Distance Modulus (DM) of 34 mag.
We chose this limit as the typical absolute magnitude of the pre-explosion outbursts is $\sim -13$ to $-14$ mag (\citealt{Smith2011MNRAS.415..773S}, \citetalias{Ofek2014ApJ...789..104O}), and the limiting apparent magnitude of modern surveys is in the 19-21 mag range (see Sect. \ref{sec:method}).

However, inspection of the information available in the literature revealed that some objects, initially classified as SNe IIn, were instead other types of transients.
In particular, \citet{Ransome2021MNRAS.506.4715R} revised the classification of a large sample of transients previously identified as SNe IIn. This allowed us to exclude Gap Transients \citep{Pastorello2019NatAs...3..676P} and SNe mis-classified as Type IIn from our sample.
Flash spectroscopy SNe \citep{Gal-Yam2014Natur.509..471G, Bruch2021ApJ...912...46B} were also removed from the sample, as they show high ionization lines (from C~III, C IV, N IV) which are rarely (if ever) observed in Type~IIn SNe. These features remain visible only for a few days, then they disappear and the object evolves as a normal Type~II SN, such as the notable case of SN 2020tlf (Jacobson-Galan et al. 2022). Instead, true SNe IIn show narrow H lines for their entire evolution, which can last even for years.
Finally, we only considered objects for which at least some pre-explosion images were available. We remark a difference between the objects of the sample: for some objects, hundreds of images are available, while for others only a few.

The final sample of 27 SNe IIn is presented in Table \ref{tab1}.
A search for pre-SN activity of one object in our sample, SN~2013gc, was already presented in \citet{Reguitti2019MNRAS.482.2750R}, who found at least one eruptive episode before the explosion in PTF data.
SN 2010jl was also included in the samples of \citetalias{Ofek2014ApJ...789..104O} and \citetalias{Bilinski2015MNRAS.450..246B}, while SNe 2019bxq and 2019fmb were included in \citetalias{Strot2021ApJ...907...99S}. Three other SNe in our sample are in common with that of \citetalias{Bilinski2015MNRAS.450..246B}: SNe 1999el, 2008fq and 2011A.
\cite{2023A&A...677A..20M} recently studied the environments of a sample of SNe IIn, and they included SNe 2016aiy and 2016cvk.
During the advancement of this research, the nearby Type IIn SN 2021foa exploded, which fullfills the requirements to be included in our sample. We conducted a dedicated analysis of this object and a search for precursors. While we did not find evidences of pre-SN activity from the archives, the Asteroid Terrestrial-impact Last Alert System\footnote{https://atlas.fallingstar.com/} \citep[ATLAS;][]{Tonry2018PASP..130f4505T} observed an early rise and a bump in the light curve two weeks before discovery, compatible with an Event A from a SN 2009ip-like object.
The individual study of SN 2021foa has been presented in \cite{Reguitti2022A&A...662L..10R}.
In Sect.~\ref{sect:results} we describe in detail the objects for which we detected pre-SN activity, while the rest of the sample will be used in Sect.~\ref{sec:discus} to estimate the rate of the precursors.

\begin{table*}
\caption{Our sample of 27 Type IIn SNe. For each object, the IAU name, celestial coordinates, redshift, host galaxy, Galactic reddening (from \citealt{Schlafly2011ApJ...737..103S}), surveys from which we collected data and approximate time interval $t_{prec}$ before the explosion between the first and last observation of the precursors or activities are reported. 
The redshifts are from the classification spectrum of each object. Objects for which we detected one or more eruptive episodes before the explosion are highlighted in bold.}
\label{tab1}
\begin{threeparttable}[b]
\begin{tabular}{llllllllll}
\hline
IAU name & R.A. & Dec. & $z$ & Host galaxy & $A_{V,MW}$ & Surveys & $t_{prec}$ (days) \\ 
\hline
	SN 1994ak	 & 	09:14:01.47	 & 	40:06:21.5	 & 	0.0085	 & 	NGC 2782	& 0.044	& - & - \\ 
	SN 1999el	 & 	20:37:17.72	 & 	66:06:11.5	 & 	0.0044	 & 	NGC 6951	& 1.020 & - & - \\ 
	SN 2002A	 & 	07:22:36.14	 & 	71:35:41.5	 & 	0.0096	 & 	UGC 3804	& 0.071 & - & - \\ 
	SN 2002fj	 & 	08:40:45.10	 & 	-04:07:38.5	 & 	0.0147	 & 	NGC 2642	& 0.060	& - & - \\ 
	SN 2008fq	 & 	20:25:06.20	 & 	-24:48:28.0	 & 	0.0106	 & 	NGC 6907/8	& 0.174	& CHASE & - \\ 
	SN 2009au	 & 	12:59:46.00	 & 	-29:36:07.5	 & 	0.00944	 & 	ESO 443-21  & 0.250	& CHASE & - \\ 
	SN 2010jl	 & 	09:42:53.33	 & 	09:29:41.8	 & 	0.0107	 & 	UGC 5189A	& 0.075	& PTF, CHASE & - \\ 
	SN 2010jp	 & 	06:16:30.63	 & 	-21:24:36.3	 & 	0.0091	 & 	NGC 2207	& 0.239	& CHASE & - \\ 
SN 2011A\tnote{1}	 & 	13:01:01.19	 & 	-14:31:34.8	 & 	0.00892	 & 	NGC 4902	& 0.137	& CHASE & - \\ 
\textbf{SN 2011fh}	 & 	12:56:14.01	 & 	-29:29:54.0	 & 	0.00803	 & 	NGC 4806    & 0.237 & CHASE & $\leq-1235$ to 0 \\ 
	SN 2011js	 & 	02:48:04.96	 & 	-13:57:51.1	 & 	0.0139	 & 	NGC 1103	& 0.065	& CHASE & - \\ 
PTF11qnf\tnote{2}	 & 	05:44:54.14	 & 	69:09:06.9	 & 	0.01428	 & 	UGC 3344	& 0.330	& PTF & - \\ 
\textbf{SN 2013gc}   &  08:07:11.88  & -28:03:26.32  & 0.003406  &  ESO 430-G20 & 1.252 & PTF, CHASE & $-1263$ to $-230$ \\ 
	SN 2013ha	 & 	06:15:49.85	 & 	66:50:19.4	 & 	0.0131	 & 	LEDA 18729  & 0.295 & PTF & - \\	
	SN 2015bf	 & 	23:24:49.03	 & 	15:16:52.0	 & 	0.01424	 & 	NGC 7653	& 0.183	& - & - \\ 
	SN 2015da	 & 	13:52:24.11	 & 	39:41:28.6	 & 	0.00667	 & 	NGC 5337	& 0.040	& PTF & - \\ 
	ASASSN-15lf	 & 	12:06:45.56	 & 	67:09:24.0	 & 	0.0083	 & 	NGC 4108	& 0.049	& PTF &	- \\ 
\textbf{SN 2016aiy}	 & 	13:08:25.39	 & 	-41:58:50.2	 & 	0.01     & 	ESO 323-G84 &	0.354 & DES & $-1070$ to $-249$ \\ 
\textbf{SN 2016cvk}	 & 	22:19:49.39	 & 	-40:40:03.2	 & 	0.0107	 & 	ESO 344-G21	&	0.035 & CHASE, DES & $-1117$ to $-319$ \\ 
SN 2016ehw\tnote{3}	 & 	08:36:37.50	 & 	73:35:04.8	 & 	0.012	 & 	PGC 24209	& 0.054	& PTF & - \\ 
	SN 2017gas	 & 	20:17:11.35	 & 	58:12:09.0	 & 	0.01	 & 	anonymous	& 1.066 & PTF, ZTF & - \\	
SN 2018lkg\tnote{4}	 & 	07:06:34.89	 & 	63:50:55.3	 & 	0.0107	 & 	UGC 3660	& 0.131	& PTF, ZTF & - \\ 
\textbf{SN 2019bxq}  & 	16:57:58.58	 & 	78:36:13.5	 & 	0.014	 & 	LEDA 84787	& 0.137 & PTF, ZTF & $-3552$ to $-18$ \\ 
	SN 2019esa	 & 	07:55:00.89	 & 	-76:24:43.1	 & 	0.00589	 & 	ESO 035-G18 & 0.497 & DES & - \\ 
\textbf{SN 2019fmb}	 &	12:26:43.68  & 	56:04:32.8   & 	0.016	 & 	LEDA 40731 	& 0.044 & ZTF & $-136$ to $-19$ \\	
	SN 2019njv	 & 	20:19:57.2	 & 	15:22:38.7	 & 	0.01458	 & 	anonymous	& 0.451	& ZTF & - \\ 
\textbf{SN 2021foa}\tnote{5} & 13:17:12.35	& -17:15:25.77 & 0.008386	& IC 863 & 0.223 & PTF, CHASE & $-20$ to $-1$ \\
\hline
\end{tabular}
\begin{tablenotes}
 \item[1] SN Impostor according to \cite{DeJaeger2015ApJ...807...63D} \item[2] Possibly an Impostor for \cite{Nyholm2017A&A...605A...6N} \item[3] Classification as SN IIn from \cite{Pan2016ATel.9129....1P,Tsvetkov2020PZ.....40....4T} 
 \item[4] Likely AGN activity or a Tidal Disruption Event, see Appendix \ref{appendix} 
 \item[5] Precursor activity observed by ATLAS
\end{tablenotes}
\end{threeparttable}
\end{table*}


\section{Methodology and Data reduction} \label{sec:method}
Because the eruptive episodes are expected to be faint, and hosted in complex environments in their host galaxies, we adopted the template-subtraction technique. 
We used a dedicated pipeline called \texttt{SNOoPY}\footnote{SNOoPy is a package for SN photometry using PSF fitting and/or template subtraction developed by E. Cappellaro at the Padova Astronomical Observatory. A package description can be found at http://sngroup.oapd.inaf.it/snoopy.html.} \citep{CappellaroSNOOPY}, that allows us to perform the alignment, transformation and PSF-matching of the template to the science images. 
The instrumental magnitudes were determined through the PSF-fit measurement performed on the template-subtracted images. For Sloan-filter images, the photometric zero points and colour terms were computed through a sequence of reference stars from the Panstarrs DR1\footnote{https://outerspace.stsci.edu/display/PANSTARRS/} \citep[PS1;][]{Chambers2016arXiv161205560C} survey in the SN field. For the few objects with $\delta< -30^{\circ}$, whose fields are not covered by the PS1 survey, we constructed a local sequence of stars from the APASS DR10 catalogue\footnote{https://www.aavso.org/apass} in the Sloan $gri$ filters. APASS DR10 was also used to calibrate the Johnson $BV$ filters frames. We calibrated the few photometric points obtained with Johnson $RI$ filters as Sloan $ri$ magnitudes. Small deviations (few hundredths of mag) in the magnitudes can arise because of this approximation. Photometric errors were estimated through artificial star experiments \footnote{In these experiments, a fake star with the same magnitude and profile of the fitted source is placed in the template-subtracted image, at a position close but not coincident with the SN position. The magnitude of the fake star in the artificial image is then measured through the same procedures described above. The dispersion of different measurements obtained from a number of iterations, with the artificial stars placed in slightly different positions, is then taken as an estimate of the instrumental magnitude error.}, also accounting for the uncertainties in the PSF-fitting procedure. Clear filter magnitudes were usually scaled to Sloan-$r$ magnitudes, because the quantum efficiency of the detector peaked at a wavelength near the response peak of that filter, while those of SN 2011fh were scaled to Johnson $V$ magnitudes.
We adopted a rather loose constraint of a Signal-to-Noise Ratio (SNR) higher than 3 (instead of 5 as used by \citetalias{Strot2021ApJ...907...99S}) for a source at the expected SN position in the template-subtracted image, to be securely detected. Nonetheless, we remark that the presence of an outburst is anyway confirmed by several detections.

We made use of reference frames (templates) without any signature of a source at the SN location down to their limiting magnitudes. These templates are then subtracted to our science frames, to remove the contamination of the host galaxy.
For the Sloan-filter images, we used the stack\footnote{https://ps1images.stsci.edu/cgi-bin/ps1cutouts} PS1 images as templates, while for the $B$ and $V$ frames we used the stack PS1-$g$ band images. For the $R$ and $I$ images, we adopted PS1-$r$ and PS1-$i$ templates, respectively. The mismatch between the Johnson and Sloan filters' response curves may introduce an additional error in the final magnitudes depending on the colour of the sources. However, this error does not influence its detectability, and only slightly changes the upper detection limits.
Finally, for southern objects we used the best-quality (in terms of seeing and exposure time) images available in the public archives as templates.
Instead, for SNe 2016cvk and 2016aiy we used aperture photometry, as a source is always visible at the SN position in the deepest images.

We searched for pre-explosion images in the public archive of the PTF\footnote{https://irsa.ipac.caltech.edu/applications/ptf/} and ZTF\footnote{https://irsa.ipac.caltech.edu/applications/ztf/} surveys, which scanned the sky with a regular cadence and have a long time coverage. Both surveys operate with the 1.2-meter Samuel Oschin Telescope at Palomar Observatory and monitor the entire northern sky ($\delta \gtrsim -30^{\circ}$) every few days. PTF operated between March 2009 and December 2012, mostly with a Johnson-$R$ filter, but also with a Sloan-$g$ filter. The standard exposure time per frame is 60 seconds, yielding a 5-sigma limiting magnitude of 20.5 and 21 mag for $R$- and $g$-band respectively.
ZTF started operations in early 2018, and with Data Release 7, all images taken until June 2021 are publicly available. ZTF routinely observed with the Sloan $g$ and $r$ filters, but occasionally observations were also performed in Sloan $i$. ZTF used a camera with a 47 square degree field of view and was able scan more than 3750 square degrees per hour to a depth of 20.5 mag \citep{Bellm2019PASP..131a8002B}. 
We also searched for images available in the archive of the CHilean Automatic Supernova sEarch survey \citep[CHASE, PI Pignata;][]{Pignata2009AIPC.1111..551P}, which operated with the Panchromatic Robotic Optical Monitoring and Polarimetry Telescopes (PROMPT; \citealt{Reichart2005NCimC..28..767R}) at the Cerro Tololo Inter-American Observatory (CTIO). The survey was primarily made with an Open filter and had a 3-sigma limiting magnitude of 19-19.5 mag.
Lastly, the Dark Energy Survey (DES\footnote{https://www.darkenergysurvey.org/}; \citealt{DES2018ApJS..239...18A}), which operates with the DECam camera mounted at the 4-m Blanco telescope at CTIO, also provided a few pre-explosion images.
The cadence of the DES survey is more relaxed compared to those dedicated to transients search, but operating with a much larger telescope it is able to reach a limit of 22.5-23 mag.
PTF monitored the fields of 11 objects in our sample before the explosion, while ZTF monitored the fields of 5 objects, CHASE 10 targets and DES 3 targets.

Along with the archives of the public surveys PTF, ZTF, DES and CHASE, we searched for pre-explosion images in the archives of major astronomical observatories worldwide. 
A drawback of the observatories archives is that the images, when available, are few and sparse in time. Nonetheless, in some cases, these images are deeper than those from the surveys, as they were taken with larger telescopes and/or longer exposure times. This allowed us to put more stringent constraints (upper limits) on the brightness of the progenitor during the quiescent phases. 
For instance, around 10 years before the explosion of SN 2011A, its progenitor was fainter than $M_R=-10.6$ mag, while in 2016 (3 years before the precursor activity we spotted) the progenitor of SN~2019fmb was dimmer than $M_g=-10.9$ mag, and finally from a deep DES image we can rule out that in 2016 the progenitor of SN 2019esa was brighter than $M_r=-10.5$ mag (see last column of Table \ref{tab1}).
The information on the facilities that provided archival data is reported in Table \ref{tab:observatories}.
The photometric tables with the magnitudes of the pre-SN detections and non-detections are available on a dedicated website\footnote{https://sngroup.oapd.inaf.it/gap.html}.


\section{Results}\label{sect:results}
In this section we describe in more detail five among the seven SNe for which we detected a precursor activity. SNe 2013gc and 2021foa were presented in \citet{Reguitti2019MNRAS.482.2750R} and \citet{Reguitti2022A&A...662L..10R}, respectively.

\subsection{SN 2011fh}
SN 2011fh was discovered by Berto Monard on 2011 August 24 \citep{Monard2011CBET.2799....1M} in the spiral galaxy NGC 4806 ($DM=32.59$ mag), at an unfiltered magnitude of 14.5. From the analysis of archive images, the object became gradually brighter from February 2011 (at 17.0 mag) to  August 2 (at 16.0 mag; this is the last observation prior to discovery).
An optical spectrum was obtained 5 days after discovery at the du Pont 2.5-m telescope at Las Campanas Observatory. The spectrum showed a blue continuum and strong emission lines from the Balmer series with multiple velocity components, typical of a Type IIn SN  \citep{Monard2011CBET.2799....1M}. 
 
The CHASE observed the field from April 2008, and followed it until March 2015. In this case we calibrated the unfiltered images to the $V$-band photometry.
The transient shows an overall rise with some fluctuations in the 3 years before the discovery. The luminosity increased with time, from a mean magnitude of 18.2 mag in 2008 to 18.0 mag in 2009 and 2010. Between January and May 2011 the object further brightened from 17.5 to 17.0 mag, and finally in Summer 2011 a major brightening marks the start of the brightest event, which is the putative Type~IIn SN explosion.

\begin{figure}
    \includegraphics[width=1.02\columnwidth]{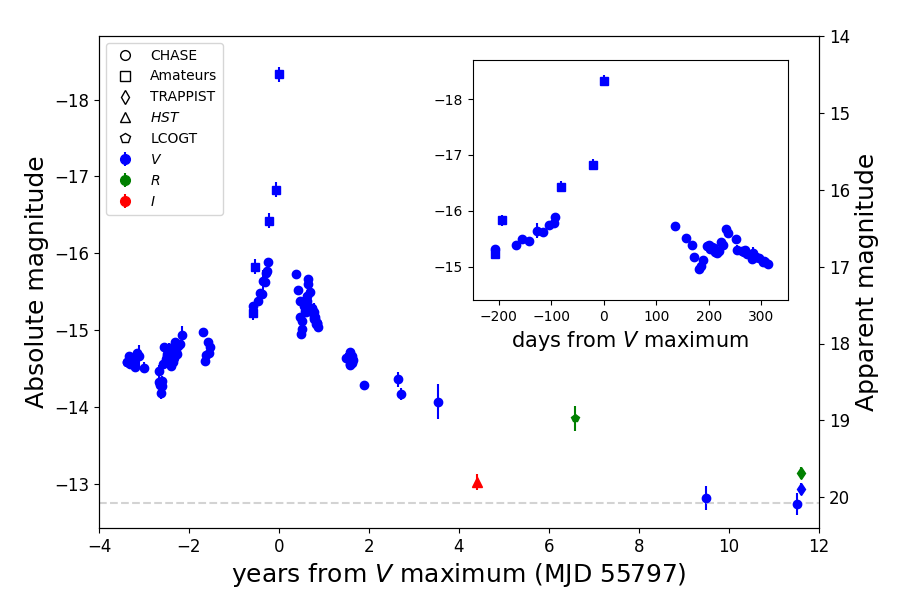}
    \caption{Pre- and post-discovery observation of SN 2011fh. Clear filter magnitudes from the CHASE Survey and from amateurs are plotted as $V$-band, a single $w$-band LCOGT observation as $R$-band and the 2016 $HST$ F814W observation as $I$-band. Observations from different instruments are plotted with different symbols. The late-time luminosity level is marked with a dashed line. A zoom on the rise to the 2011 event and the first year of decline, during which a second, smaller peak is visible at +250 days, is plotted in the blow-up window.}
    \label{fig:11fh}
    \end{figure}

The CHASE survey resumed observing SN 2011fh on 7 January 2012, when the transient had already dimmed to mag 17.1, continuing to fade with a slower decline rate. In fact, in March 2015 (nearly 4 years after the explosion) the object was still recovered at $V$=18.8 mag, slightly fainter than the luminosity level the source was spotted at in 2008. 
In 2016 (5 years after discovery) the Hubble Space Telescope ($HST$) observed the field of NGC 4806 with the WFC3 camera in the F336W and F814W filters (P.I. Filippenko). A bright source was clearly visible within 0.1" from the location of SN 2011fh. From that image, we measured $F814W=19.8\pm0.1$ mag, indicating a clear flattening of the light curve. The transient was still detectable in March 2018 from an archival white-filter LCOGT image at $R\simeq19.0\pm0.2$ mag, and even more recently in February 2021 and March 2023 from an unfiltered PROMPT image at $V=20.0\pm0.16$ and $V=20.1\pm0.14$ mag, respectively. All our data are shown in Fig. \ref{fig:11fh}.

In their paper, \cite{Pessi2022ApJ...928..138P} studied SN 2011fh in detail. Their photometric data cover the 2007-2013 period, and their $r$-band light curve is very similar to our $V$-band light curve. They concluded that SN 2011fh shares common features with the Type IIn SN 2009ip \citep{Mauerhan2013MNRAS.430.1801M, Pastorello2013ApJ...767....1P, Margutti2014ApJ...780...21M, Smith2014MNRAS.438.1191S}. They attribute the $HST$ detection at +5 years to still on-going ejecta-CSM interaction or to an eruptive phase. A MUSE spectrum at the Very Large Telescope (VLT) of the field of SN 2011fh indeed shows signs of interaction at nearly +4 years.
Other members of the SN 2009ip-like family of objects have recently been confirmed to be genuine terminal SNe events, as they continued to fade below the progenitor level after the brightest event \citep{Smith2022MNRAS.515...71S, Jencson2022ApJ...935L..33J, Brennan2022A&A...664L..18B}. 
\cite{Fransson2022A&A...666A..79F} propose multiple scenarios to explain SN 2009ip-like events: a pulsational pair instability
SN \citep{Woosley2007Natur.450..390W}, a large mass eruption followed by a normal SN explosion from a lower mass ($\lesssim$20 \msun) progenitor, or a merger of a massive star with a compact object.
In contrast, SN~2011fh is still visible after more than 10 years, although it is slightly fainter than 3 years before the 2011 event.

\begin{figure}
    \includegraphics[width=1.02\columnwidth]{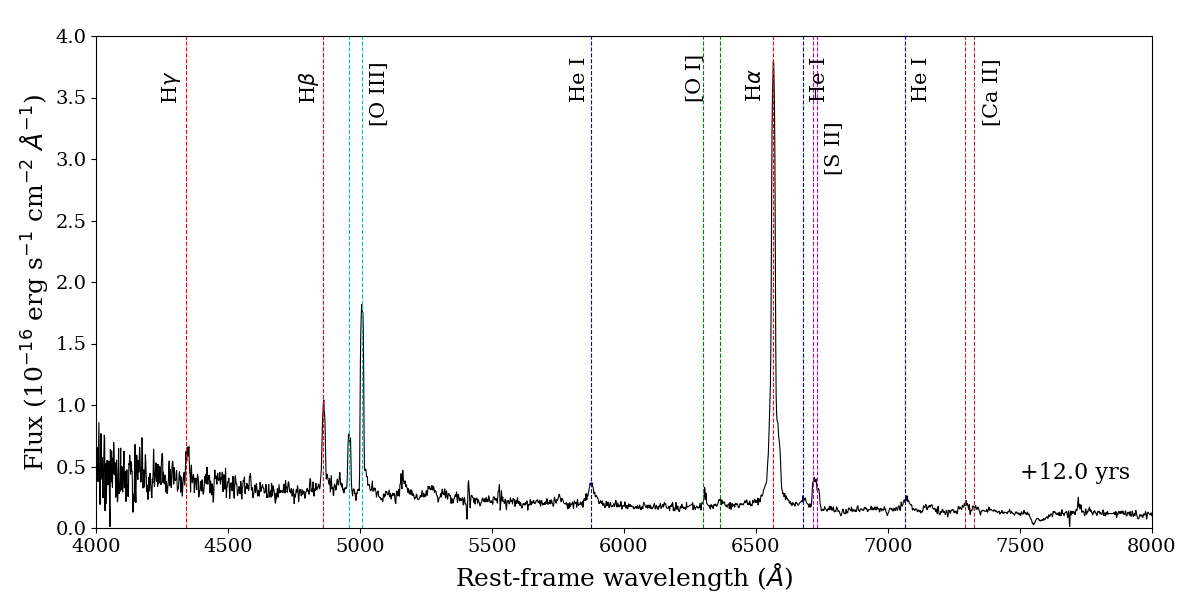}
    \caption{A spectrum of SN 2011fh taken with the 6.5-m Clay telescope+LDSS3 12 years after maximum light. The principal identified emission lines are marked.}
    \label{fig:11fh_spectrum}
    \end{figure}

On 10 August 2023 (12 years after maximum) we obtained a spectrum of SN 2011fh with the 6.5-m Clay telescope, which was reduced, extracted and calibrated using routine \texttt{IRAF} procedures. The final spectrum is shown in Figure \ref{fig:11fh_spectrum}.
Overall, the spectrum does not show a significant evolution from the +4~years MUSE spectrum published by \cite{Pessi2022ApJ...928..138P}, in accordance with the extremely slow evolution of the light curve at very late phases.
He I lines are present, with a full width at half maximum (FWHM) velocity comparable to that of the broader component in the H$\alpha$ line (1500-2000 \kms).
The FWHM of the He I lines is 1700 \kms, compatible with the velocity of a fast outflow from an early Wolf-Rayet star \citep{Smith2017hsn..book..403S}.
The [O I] and [Ca~II] doublets are absent or very faint, suggesting that the progenitor star survived the 2011 event, making SN 2011fh a luminous SN impostor.

\subsection{SN 2016aiy}
The discovery of SN 2016aiy (ASASSN-16bw) was made by the All Sky Automated Survey for SuperNovae \citep[ASAS-SN;][]{Shappee2014AAS...22323603S} on 2016 February 17 \citep{Brimacombe2016TNSTR.123....1B}, in the galaxy ESO 323-G084 ($DM=33.07$ mag), at $V=16.9$ mag. However, a detection from ASAS-SN was obtained 2 days earlier, at $V=17.2$ mag. The last non-detection is on February 11, at $V>18.2$ mag.
SN 2016aiy was classified as a SN IIn on 2016 February 24 by the PESSTO Collaboration \citep{Taubenberger2016TNSCR.150....1T}.

\begin{figure}
\includegraphics[width=1.02\columnwidth]{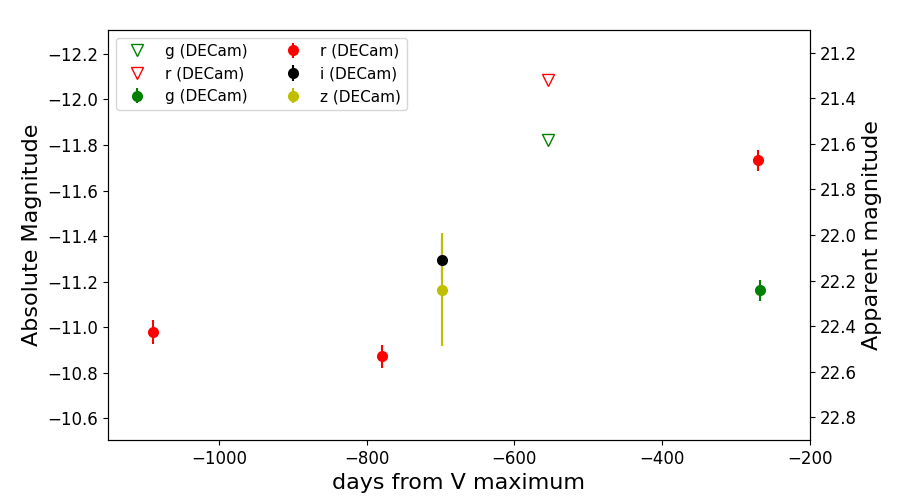}
\caption{Pre-discovery $griz$ observations of SN 2016aiy from the DES survey. The phases are relative to the $V$-band maximum (Reguitti et al., in prep.).}
\label{fig:16aiy}
\end{figure}

The field of SN 2016aiy was monitored before the explosion only by the DES Survey between March 2013 and June 2015.
We have only four epochs of sparse $griz$ photometry, and in three of them we measured a source with magnitudes between 21.6 and 22.6 mag. In particular, the object was detected thrice in the $r$-band, at the extremes of the monitoring temporal window, at $M_r$ between $-10.8$ and $-11.8$ mag; see Fig. \ref{fig:16aiy}).

We conclude that we are observing the progenitor in an active state or in outburst, because the source is variable and a $-11$ mag star in quiescence is less plausible.
We can rule out a variable stellar source within a luminous stellar cluster, as in very late time (+3 yrs) and profound images from the NTT telescope, we do not detect any source brighter than 22.8 mag, an upper limit which is fainter than the pre-SN detections (Reguitti et al., in prep.).

The post-explosion light curves and spectral evolution of SN 2016aiy, along with their connection to the observed pre-SN activity will be presented in a forthcoming paper (Reguitti et al., in preparation).

\subsection{SN 2016cvk}
SN 2016cvk was discovered by the Backyard Observatory Supernova Search (BOSS) team on 2016 June 12 \citep{Parker2016TNSTR.422....1P} in the galaxy ESO-344-G21 ($DM=33.21$ mag), at an unfiltered magnitude of 17.6. The last non-detection is on 2016 June 3, with a limiting magnitude of 18.0.
The ASAS-SN Survey discovered a second transient at the location of SN 2016cvk on 2016 August 31, at $V\sim$16.2 mag, 
identified with a survey name of ASASSN-16jt \citep{Brimacombe2016ATel.9439....1B}. A non-detection ($V>17.8$ mag) 5 days before discovery was also reported by \cite{Brimacombe2016ATel.9439....1B}.
The position of the new transient was within 1" from the position reported in June 2016. For this reason, we are confident that the location of the two transients is coincident, and that the event observed in June 2016 was a previous outburst of an object that later exploded as a SN (see also \citealt{Brown2016ATel.9445....1B}).
The classification spectrum of ASASSN-16jt/SN 2016cvk was taken 3 months later, on September 6, by the PESSTO Collaboration \citep{Smartt2015A&A...579A..40S} with the 3.58-m ESO-NTT telescope. The new object was classified as a peculiar SN IIn \citep{Bersier2016TNSCR.650....1B}.
However, a previous spectrum was taken on June 18 (during the outburst) at the du Pont 2.5-m telescope \citep{Brown2016ATel.9445....1B}, which is similar to that of SN 2009ip before its peak in September 2012. This, and the photometric evolution observed during Summer 2016, make SN 2016cvk a promising SN 2009ip-like candidate. An extended analysis of the post-explosion light curve and spectra evolution will be presented in a future paper (Matilainen et al., in preparation).

\begin{figure}
\includegraphics[width=1.02\columnwidth]{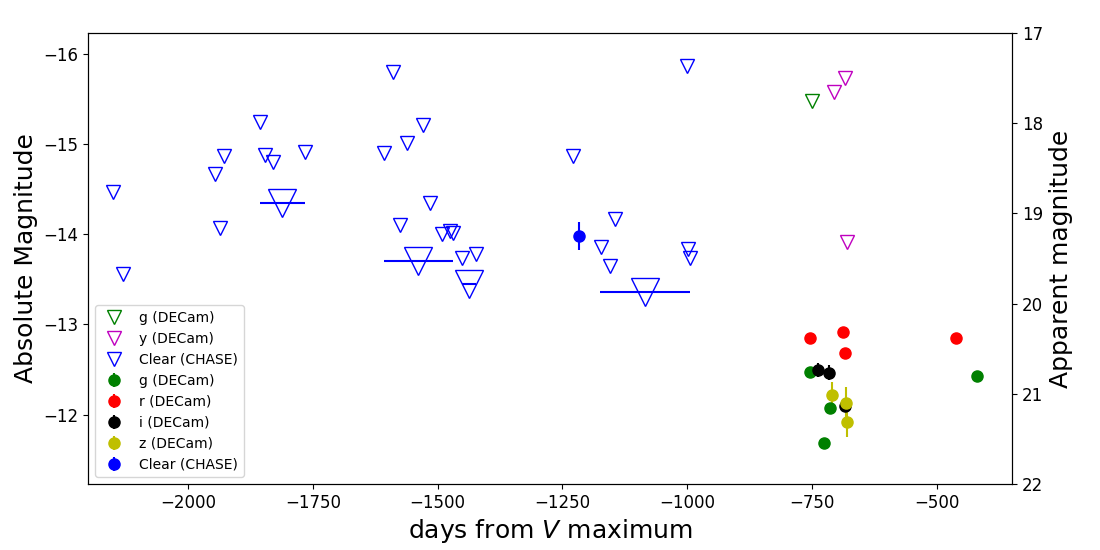}
\caption{Pre--discovery light curves of SN 2016cvk. The clear filter observations from the CHASE are in blue, the other points are with Sloan filters from the DECam instrument. The open inverted triangles are upper limits, the filled circles are detections. The larger triangles with an horizontal error bar indicate deeper upper limits derived from stacked frames within the temporal bar. The phases are relative to the $V$-band maximum (Matilainen et al., in prep.).}
\label{fig:16cvk}
\end{figure}

The CHASE project monitored the field of SN 2016cvk between October 2010 and December 2013. We find no further detections in those images, apart for a single detection on 2013 May 12 at $19.25\pm0.16$ mag (calibrated as Sloan-$r$). At the assumed distance of ESO-344-G21, this photometric point gives an absolute magnitude of $M_r\sim-14.0$ mag. Starting from November 2013, the DES survey targeted the same region of sky in Sloan-$grizy$, until July 2015. Thanks to the large diameter of the telescope, DES can provide much deeper photometry. Indeed, in all but one image, we detect a faint source at the SN location (between 20.3 and 21.5 mag) in the $griz$ filters. The pre-discovery absolute light curve is in Fig. \ref{fig:16cvk}.

\subsection{SN 2019bxq}
SN 2019bxq (ZTF19aamkmxv; PS19ahx) was formally discovered by the ZTF survey on 15 March 2019 \citep{Nordin2019TNSTR.404....1N}, in the galaxy LEDA 84787. According to the NASA Extragalactic Database (NED\footnote{https://ned.ipac.caltech.edu/}), the host is an Elliptical Star-Forming galaxy. The classification spectrum, taken at the Palomar 5-meter Hale telescope about one month after the discovery, shows numerous narrow emission lines, including those from the Balmer series, over a still blue continuum, making SN 2019bxq a Type IIn SN \citep{Fremling2019TNSCR.747....1F}. The host redshift reported by NED, $z$=0.01421, is consistent with that obtained from the classification spectrum ($z$=0.014). Using the former $z$ estimate, we derive a kinematic distance modulus $DM=33.83$ mag.
The adopted Galactic reddening is $A_{B,MW}=0.181$ mag \citep{Schlafly2011ApJ...737..103S}; using the \cite{Cardelli1989ApJ...345..245C} extinction law, we derive a reddening in the $R$-band of $A_{R,MW}=0.11$ mag.

\begin{figure*}
\includegraphics[width=2.05\columnwidth]{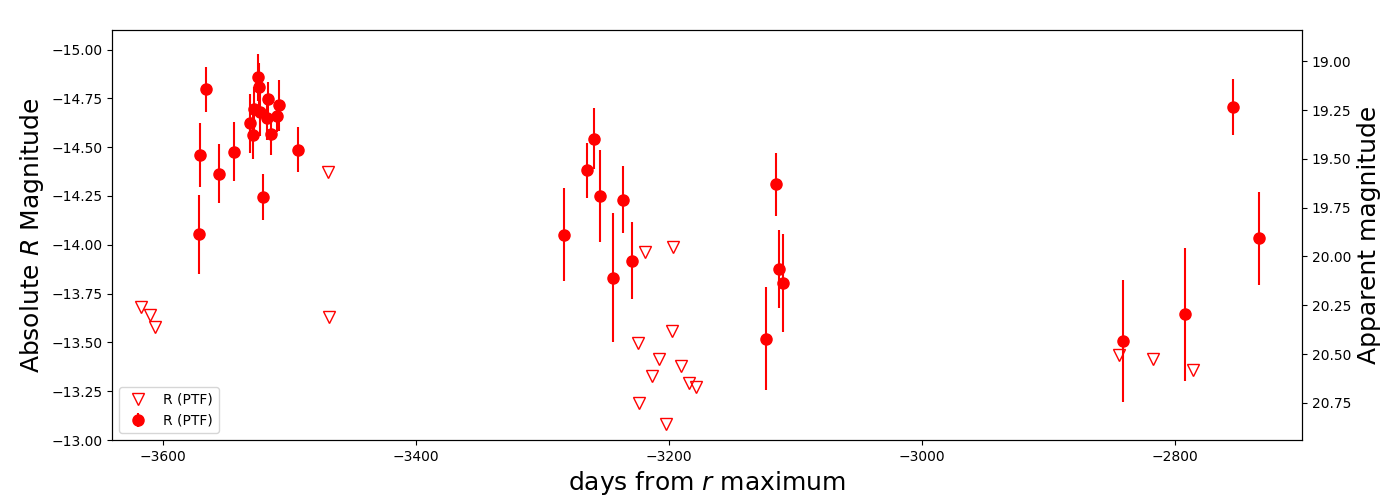}
\includegraphics[width=2.05\columnwidth]{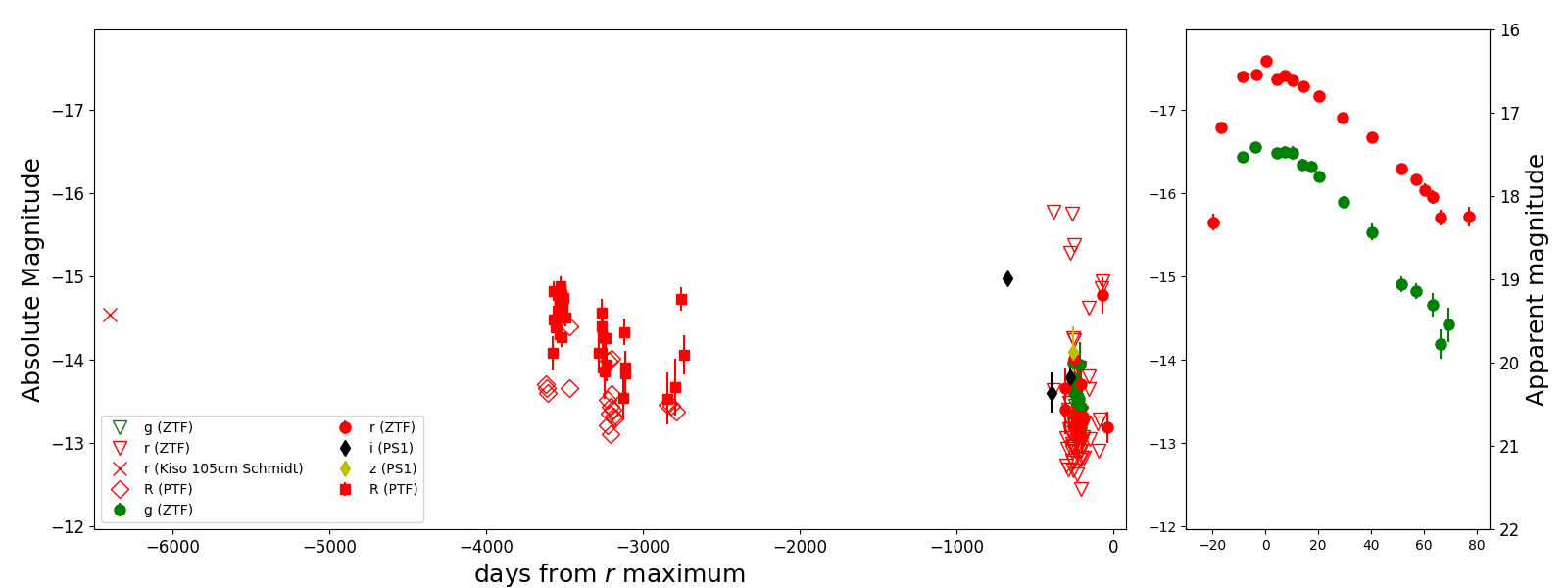}
\caption{{\it Top:} Pre-explosion $R$-band absolute light curve of SN 2019bxq from the PTF survey. The data span the 2009-2011 period. The magnitudes are corrected for the adopted distance modulus and Galactic reddening. The phase on the x-axis is from the SN $r$-band maximum ($MJD=58577$). The open inverted triangles mark upper limits (i.e. non-detections). {\it Bottom-left:} Same as above, but including all the available data (PTF+ZTF+sparse archive data). Observations from different instruments are plotted with different symbols. The phases are relative to the brightest $r$-band point in the SN light curve. {\it Bottom-right:} Post-explosion $g$- and $r$-band light curves of SN 2019bxq from the ZTF survey. The SN light curves are shown to highlight the short duration of the SN event.}
\label{fig:19bxq}
\end{figure*}

SN 2019bxq was part of the \citetalias{Strot2021ApJ...907...99S} sample, and that study found precursor activity in the ZTF images. Furthermore, we observed multiple eruptive episodes in the PTF data, from 2009 to 2011. As shown in Fig. \ref{fig:19bxq}, there are multiple detections, that can be divided into four main eruptive events. 
The first event began 3571 days ($\sim$9.8 years) before the SN maximum light, and lasted around 3 months. This was the best sampled outburst, and from the brightest detection ($R=19.08\pm0.12$ mag) we can infer a maximum absolute magnitude of the outburst of $M_{R,max}=-14.8$ mag (see also Fig. \ref{fig:19bxq}). After the seasonal gap, the object became visible again. We detected the object in outburst for the second time during the luminosity rise, reaching a similar maximum absolute magnitude of $M_{R,max}=-14.5$ mag, before declining. For a few weeks the transient was not detected above the limiting magnitude threshold of our images. Later, the object was newly detected during its third outburst. Finally, after a long interruption due to the Solar conjunction, we detected the object during a fourth outburst. The last points were obtained before the end of the PTF survey operations.
Finally, we also recovered sparse detections from the ZTF data in the final year before the explosion, already spotted by \citetalias{Strot2021ApJ...907...99S}.

\subsection{SN 2019fmb}
SN 2019fmb (ZTF19aavyvbn; ATLAS19bbhu; PS19ahl) was discovered by the PS1 survey on 12 May 2019 \citep{Chambers2019TNSTR.796....1C}, in the galaxy LEDA 40731. The classification spectrum, taken at the 3.5m APO telescope about 6 months after discovery, shows narrow H$\alpha$ emission over a continuum that is still quite blue. For this reason, SN 2019fmb was classified as a Type~IIn SN \citep{Graham2020TNSCR.603....1G}. From the H$\alpha$ emission a redshift $z=0.016$ is deduced, and thus we adopt $DM=34.09$ mag. The Galactic reddening is negligible, at $A_{r,MW}=0.036$ mag.

\begin{figure*}
\includegraphics[width=2.1\columnwidth]{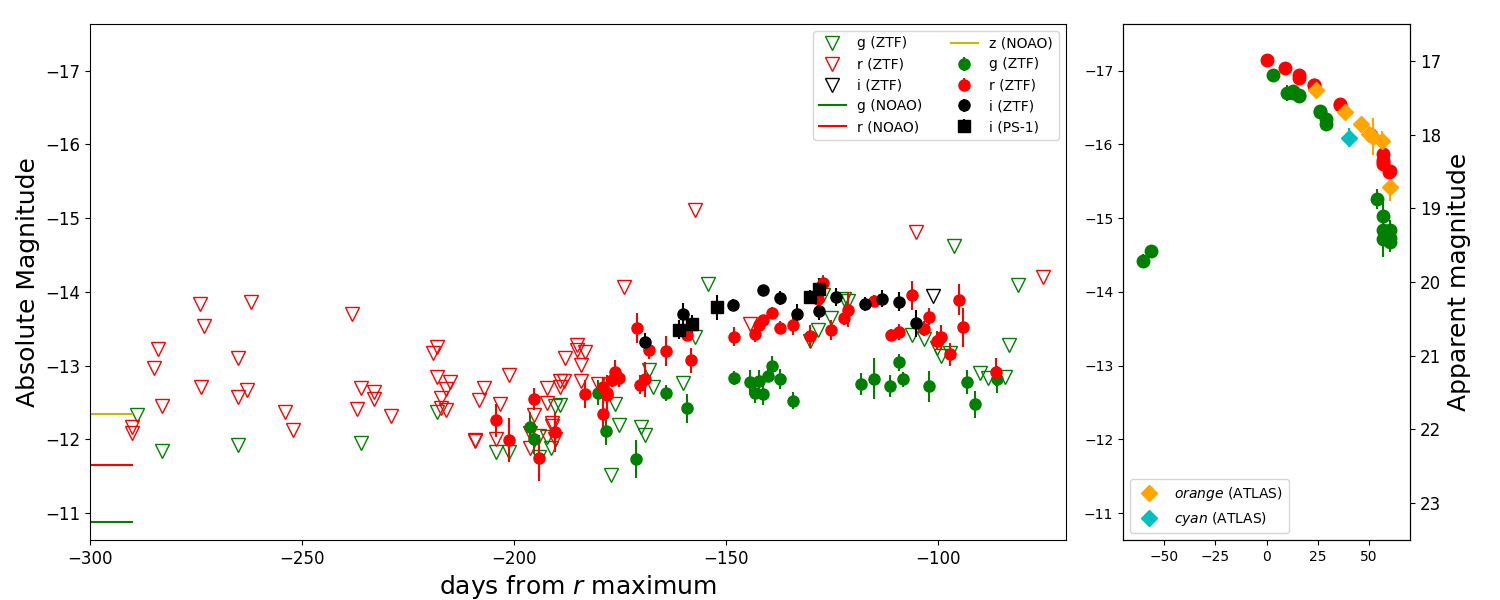}
\caption{
{\it Left:} Pre- and post-discovery (MJD = 58615) Sloan-$gri$ detections from the ZTF and PS-1 surveys of SN 2019fmb. Observations from different instruments are plotted with different symbols. The phases are relative to the brightest $r$-band point in the SN light curve. The deepest upper limits of the object in $g$-, $r$- and $z$-band filters from NOAO archival data of 2016 are marked with horizontal lines on the bottom-left corner. {\it Right:} Light curve of SN 2019fmb, with the detections by ATLAS also reported. As for SN 2019bxq, the SN has a short duration.}
\label{fig:19fmb}
\end{figure*}

Analysing the ZTF data of SN 2019fmb, we found numerous faint detections after the discovery epoch, with apparent magnitudes between 20.0 and 21.5 mag, corresponding to absolute magnitudes ranging from $-12.5$ to $-14.0$ mag. Pre-discovery detections were obtained thanks to the PS1 survey in the $g$, $r$ and also $i$ bands. The absolute magnitude of the source in this phase suggested it was most likely an SN impostor. We observed a 3-months lasting pre-SN activity, with the $r$-band luminosity gently increasing to a shallow maximum at around one month after the  discovery, and then the outburst luminosity slightly faded. About 100 days after, we detected the object twice in the $g$ band, with the source becoming much brighter ($M_g\sim-14.5$ mag) than during the previous activity. We argue that these detections mark the light curve rise to the maximum light (which was not observed) soon after the SN explosion. The SN luminosity decline was partially followed by the ATLAS survey (Fig. \ref{fig:19fmb}, right panel).

The light curve hump observed before the explosion of SN~2019fmb is reminiscent of those observed before the explosion of other SNe IIn, specifically SN 2009ip-like events (e.g. SNe 2016bdu, \citealt{Pastorello2018MNRAS.474..197P}; LSQ13zm, \citealt{Tartaglia2016MNRAS.459.1039T}; 2016jbu, \citealt{Brennan2022MNRAS.513.5642B}; 2021foa, \citealt{Reguitti2022A&A...662L..10R}).
SN~2019fmb is also included in the sample of \citetalias{Strot2021ApJ...907...99S}, and they found a similar photometric behaviour. The corresponding light curves are plotted on top of Fig. \ref{fig:19fmb}. 

We also found a bunch of older ‘Mosaic3+90prime' Surveys images from the NOAO Archive taken in 2016 and 2017. While the transient is not detected in this images, we can establish an upper limit to the $g$-band quiescent progenitor magnitude, with $M_g>-11$ mag (Fig. \ref{fig:19fmb}, bottom left).

\section{Discussion} \label{sec:discus}

Among the objects considered in our sample, SNe 2013gc, 2016aiy, 2016cvk, 2019fmb, and 2021foa present one pre-SN outburst or prolonged activity each, while SN 2019bxq shows four events, and SN 2011fh reveals a slow years-long rise, noted also by \cite{Pessi2022ApJ...928..138P}. Hence, we observed 10 luminous events before the 27 SNe in our sample. 

Unlike \citetalias{Strot2021ApJ...907...99S}, the precursors in this work are detected not only close to the SN explosion, but also at longer times before (up to 10 years prior to the SN explosion in the case of SN~2019bxq).
Unstable ignition of advanced nuclear burnings in the core of massive stars is sometimes invoked as a powering mechanisms for SN impostor events \citep{Arnett2011ApJ...741...33A, Quataert2012MNRAS.423L..92Q, Smith2014ApJ...785...82S}. Specifically, \cite{Shiode2014ApJ...780...96S} show that in stars with $M\lesssim20$ \msun the Ne-burn can start 1 to 10 years before the SN explosion, which is broadly compatible with the precursor activity observed before SNe 2019bxq, 2016aiy and 2016cvk.
\cite{Wu2021ApJ...906....3W, Wu2022ApJ...930..119W} updated the models by \cite{Fuller2017MNRAS.470.1642F} and \cite{Fuller2018MNRAS.476.1853F}, finding that the wave-driven heating is not strong enough to trigger mass loss in H-rich stars, hence this mechanism is not able to explain the precursor of SNe IIn.

\subsection{Precursor characterisation} \label{sect:prec_ch}
\begin{figure}
\includegraphics[width=1.05\columnwidth]{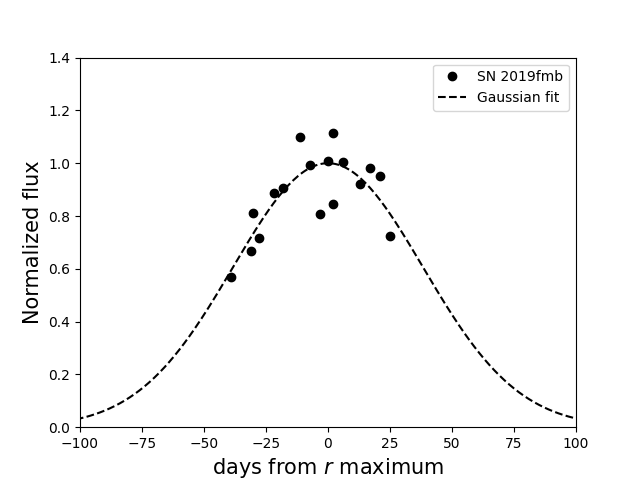}
\caption{The Gaussian fit (dashed line) over the $r$-band light curve of the pre-SN outburst before SN 2019fmb. The magnitudes of the data-points were converted into fluxes, then normalised with respect to the peak of the Gaussian fit and centred on the epoch of the maximum.}
\label{fig:points}
\end{figure}

Here we try to characterise the precursors of SNe IIn we found by analysing their observational properties, such as duration, luminosity and colour, looking for common features.
In general, the precursor activities we detected are not well sampled, therefore we need to make some strong assumptions about the shape of the burst light curve in order to have a first order characterisation of them. With the purpose of limiting the complexity of the model, we assume that the flux evolution of each burst can be described with a Gaussian function with four free parameters: the amplitude ($a_p$), the width ($\sigma_p$), the epoch of the maximum and the background level.
To increase the statistics, we collected the published magnitudes of other outbursts before Type IIn SNe that do not belong to our sample, namely SNe 2015bh \citep{EliasRosa2016MNRAS.463.3894E,Thone2017A&A...599A.129T}, 2016bdu \citep{Pastorello2018MNRAS.474..197P}, 2018cnf \citep{Pastorello2019A&A...628A..93P}, LSQ13zm \citep{Tartaglia2016MNRAS.459.1039T}, 2010mc \citep{Ofek2013Natur.494...65O}, 2011ht \citep{Mauerhan2013MNRAS.431.2599M, Fraser2013ApJ...779L...8F}.
Among them, SNe 2016bdu and 2018cnf show two pre-SN bright events, that are treated as two separate outbursts.
For all the precursors, the $r$- or $R$-band filter has the largest number of detections, thus we focused our analysis on this band. We converted the apparent magnitudes to luminosity by adopting the distances and Galactic reddenings reported in Table \ref{tab1} or in the relative papers. For each outburst, we performed the previously mentioned Gaussian fit over the luminosities with the \texttt{curve\_fit} tool in \textit{Python}. Upper limits were also used to constrain the fits. An example of a Gaussian fit on the data points of an outburst is shown in Fig. \ref{fig:points}. 

For each Gaussian best-fit, we calculated the area $A_p=a_p*\sqrt{2\pi\sigma_p^2}$, which is the most characteristic parameter for statistical purposes. In physical terms, the area would be proportional to the total amount of energy released during the outburst. 
We can also define a mean luminosity during the outbursts by dividing the area of the Gaussian ($a_p\sigma_p\sqrt{2\pi}$) by the maximum considered temporal length ($6\sigma_p$), obtaining $\overline{L_p}=a_p*\sqrt{2\pi}/6\approx0.42*a_p$, which is $\approx$0.95 mag fainter than the peak magnitude.

The background level was set to 1 million \lsun\,(equal to $M_r=-9.8$ mag). This value was chosen as it is close to the estimated luminosity of the quiescent progenitors of the type IIn SNe 2005gl \citep{Gal2007ApJ...656..372G, Gal2009Natur.458..865G} and 2009ip \citep{Foley2011ApJ...732...32F} identified in \textit{HST} archival images as candidates LBVs.

The observed brightest absolute magnitudes of the precursors range between $-11.5$ and $-14.8$ mag, with an average of $-13.7\pm1.5$ mag. 
The faintest detection of a precursor is for SN 2016aiy at $M_r=-10.9$ mag, thanks to the deepness of the DES survey images.
Therefore we do not find any precursor brighter than $M_r$ $-15$ mag (excluding SN 2018lkg, which we assume not to be a SN IIn, see Appendix \ref{appendix}). The range in magnitude of the single detections is $-10.9$ to $-14.8$, similar to but narrower than that of the events identified by \citetalias{Strot2021ApJ...907...99S} (from $-12$ to $-17$ mag).
Our sample is smaller than that of  \citetalias{Strot2021ApJ...907...99S}, this may explain why we do not detect very bright outbursts. At the same time, we may also miss faint events because we did not stacked our data.

We searched for a correlation between $\sigma_p$ and $a_p$ of the gaussians which can be interpreted as a proxy for the duration and peak brightness of the burst, respectively, but we did not find any (Pearson correlation test: r=0.11 and p-value=0.69).

\begin{figure}
\includegraphics[width=1.02\columnwidth]{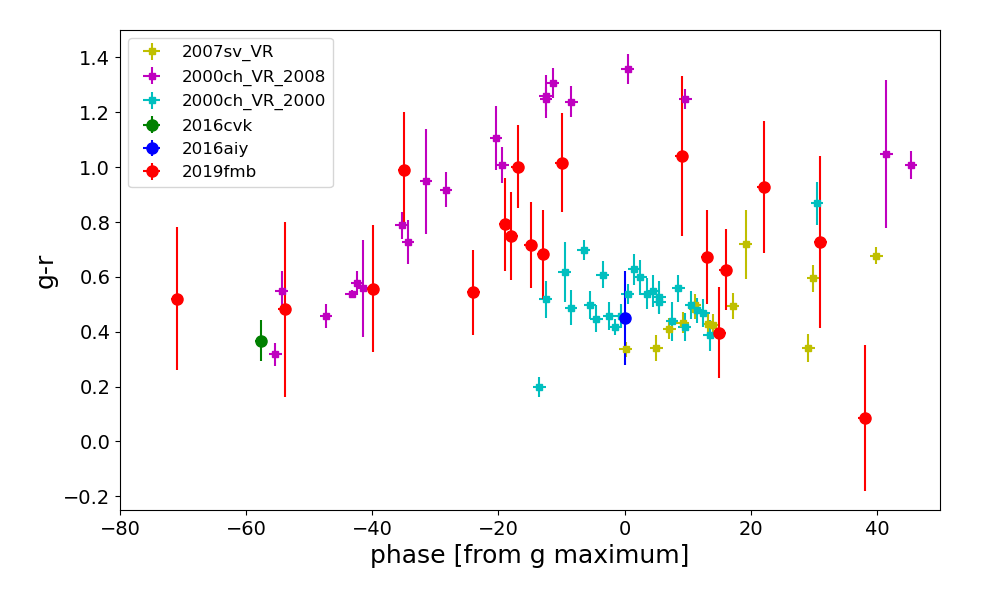}
\caption{The $g-r$ colour of the 3 pre-SN outbursts or activities for which multi-band photometry is available. The phases are relative to the maximum in the $r$-band, derived from a Gaussian fit over the data-points. For comparison, the $V-R$ colour of the 2000 and 2008 outbursts of the known LBV AT 2000ch \citep{Pastorello2010MNRAS.408..181P} and of the SN impostor SN 2007sv \citep{Tartaglia2015MNRAS.447..117T} are also shown. The $g-r$ colour of the outbursts is red, around $g-r\sim$0.5 mag.}
\label{fig:colors}
\end{figure}

For three objects, pre-explosion outbursts were detected in more than one filter, thus allowing us to construct their colour curves.
Because most of the outbursts were observed in recent years, multi-band observations are available in the Sloan-$g$, $r$ and $i$ filters. This allows us to calculate the evolution of the \textit{g-r} and \textit{r-i} colours.
We considered two observations in different filters obtained within less than 1 day from each other as contemporary.
In Fig. \ref{fig:colors} we plot the \textit{g-r} colour light curves of the outbursts of SNe 2016aiy, 2016cvk and 2019fmb with respect to the epoch of maximum of the outbursts. The maximum light epoch of each outburst was estimated using gaussian fits.

The $g-r$ colour of the 3 outbursts are concentrated in the range 0.5 to 1.0 mag. This red colour is compatible with that observed during the giant eruptions of LBV stars, whose spectra shift from those of B-types to cooler F-types \citep{Humphreys1994PASP..106.1025H}. For example, 
during the 2014-2017 eruption of the LBV R40, it had $B-V\approx$0.6 mag \citep{Campagnolo2018A&A...613A..33C}, corresponding approximately to $g-r\approx$0.5 mag \citep[using the relations of][]{Jester2005AJ....130..873J}.
In Fig. \ref{fig:colors}, the \textit{g-r} colours of the outbursts are compared with the \textit{V-R} colour of the SN impostor SN 2007sv \citep{Tartaglia2015MNRAS.447..117T}, together with the 2000 and 2008 bright events of the LBV known as AT 2000ch \citep{Wagner2004PASP..116..326W, Pastorello2010MNRAS.408..181P}. The colours are similar, especially with those of the 2000 event.
Focusing on the $g-r$ colour curve of the precursor of SN 2019fmb, despite its relatively large error bars, we note that the colour shows an evolution during the outburst, passing from blue ($g-r\sim$0.5 mag) to red (1.0 mag) from the begin to the maximum, and then back to blue at the end of the burst.

\begin{figure}
\includegraphics[width=1.0\columnwidth]{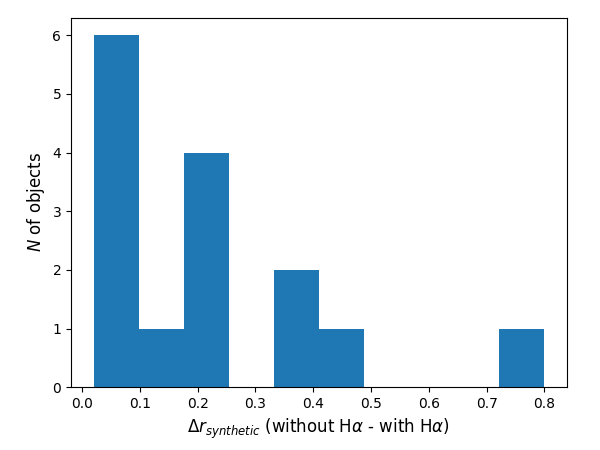}
\caption{Histogram of the distribution of the difference in the $r$-band synthetic photometry, with and without the H$\alpha$ contribution, of the classification spectra for a sample of SN impostors and LBV eruptions.}
\label{fig:hist}
\end{figure}

The spectra of SN impostors and LBV eruptions are usually characterised by strong and narrow emission lines, with H$\alpha$ being predominant. As the maximum of the response curve of the $r$ filter is close to the wavelength of H$\alpha$, its flux may affect the \textit{g-r} colour.
To establish if the H$\alpha$ flux contribution can account for the red colours of the pre-SN outbursts, we collected the classification spectra of all the objects classified as SN impostors or LBV eruptions in the TNS and WISEReP databases (15 objects in total), and manually removed the H$\alpha$ line in each of them. Then, we calculated the $r$-band synthetic photometry with the IRAF \texttt{synphot} tool, both with and without including H$\alpha$, and calculated the difference in magnitudes. In most cases, the difference is less than 0.3 mag, for three objects is around 0.4 mag and just one at 0.8 mag. The histogram of the differences distribution is shown in Fig. \ref{fig:hist}. The small differences indicate that the H$\alpha$ emission can only partially justify the observed red $g-r$ colours of the outbursts, hence this is mostly due to a cooler continuum temperature, as seen in the LBV eruptions.

For the precursor of SN 2019fmb it is possible to obtain the \textit{r-i} and \textit{g-i} colours as well. Its mean \textit{r-i} colour is 0.5 mag, while the \textit{g-i} colour is around 1.1 mag. The \textit{g-i} value is higher than the 2008 outburst of AT 2000ch (which rose from 0.4 to 0.9 mag), and is the same found by \citetalias{Strot2021ApJ...907...99S}, which also agrees on the low effective temperatures of the precursors ($\sim$4300 K for SN 2019fmb).

To summarise, the outbursts we found are rather luminous (with absolute magnitudes in the $-11$ to $-15$ mag range), red in colour (with $g-r$ from 0.5 to 1.0 mag) likely due to cool photospheric temperatures, analogous to that observed during LBV eruptions.

\subsection{Connection between pre-SN outbursts and SN light curves}
We investigated a possible impact that the pre-explosion activity could have on the properties of the SNe IIn after the explosion. For SNe 2019bxq and 2019fmb, we collected the publicly released $gr$ light curves from the ZTF survey,
available\footnote{https://alerce.online/object/ZTF19aavyvbn\\https://alerce.online/object/ZTF19aamkmxv} from the \textit{ALeRCE} broker \citep{Forster2021AJ....161..242F}. For SN~2011fh, we used the images from the CHASE survey to construct also the post-explosion light curve.
The light curves of SNe 2016ehw, 2017gas and 2019esa are retrieved from the Gaia Photometric Science Alerts\footnote{http://gsaweb.ast.cam.ac.uk/alerts/alert/}. Finally for SNe 2008fq, 2009au, 2010jl, 2010jp and 2015da they are collected from the respective research papers \citep{Taddia2013A&A...555A..10T, Rodriguez2020MNRAS.494.5882R, Zhang2012AJ....144..131Z, Smith2012MNRAS.420.1135S, Tartaglia2020A&A...635A..39T}.

The interaction between the SN ejecta and the material ejected during the precursor activity should modify the evolution of the light curve of the SNe, as the thermalisation of the kinetic energy from the ejecta into radiation through the interaction with the CSM should provide an additional powering source \citep{Graham2014ApJ...787..163G, Nyholm2017A&A...605A...6N}. Therefore, to investigate if the precursor activity has a significant impact on the post explosion behaviour, we measured the decline rates of the post-explosion light curves of the SNe in our sample during the first 65 days post explosion. Assuming that the material ejected during the outburst episodes has a maximum velocity of $\sim$1000 \kms and that SN ejecta expand at velocity higher than $\sim$10000 \kms, they should reach the material generated in correspondence to the occurred precursor activity, as far as 65/0.1=650 days before the explosion. As visible in Figures \ref{fig:11fh} and from \ref{fig:16aiy} to \ref{fig:19fmb}, in all the SNe of our sample that show precursor activity, the last episode occurs closer to the explosion time frame than the previously mentioned time span. Therefore, if the precursor activity has a significant impact on the SNe light curve decline rate, in principle it should be detectable in our sample.

\begin{figure}
\includegraphics[width=1.0\columnwidth]{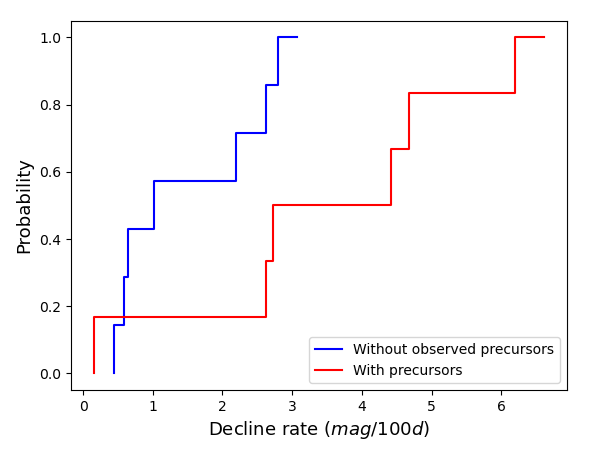}
\caption{Cumulative distributions of the decline rates of the SNe IIn in our sample with (red) and without (blue) observed precursor activity. The two distributions are clearly separated, with SNe which showed precursor events declining faster after the explosion.}
\label{fig:cdf}
\end{figure}

In order to test the latter, we constructed the cumulative distributions of the decline rates for the SNe that did and did not show precursor activity in our data (see Fig. \ref{fig:cdf}). The Kolmogorov–Smirnov test over the two cumulative distributions gives a statistic $D=0.61$ and a p-value of 0.09, while the Anderson-Darling test provides a statistic $A=2.18$ with a significance level of 0.04. The latter indicates that there is a marginally significant difference between the two distributions, but a large sample is clearly necessary to confirm or discard such result. Anyway, the occurrence of pre-SN outbursts does not seem to have a strong influence on the SNe after the explosion. \citetalias{Ofek2014ApJ...789..104O} and \citetalias{Strot2021ApJ...907...99S} reached a similar conclusion through  different, more direct, probes. 
On the other hand, if this difference is confirmed, it is somewhat unexpected that the SNe for which we found signatures of precursor events are in general the ones with a faster decline after maximum. This is because the CSM expected to be ejected during the pre-SN outbursts should later interact with the SN ejecta, converting their kinetic energy into radiation and powering the light curve. This interaction would make SNe IIn more luminous and, if the CSM is massive enough, possibly slower  declining than non-interacting SNe. 
However, a possible explanation for the rapid decline of those SNe with observed precursors is the rapid formation of a cool dense shell (CDS) of dust inside the ejecta \citep{Smith2008ApJ...680..568S}, when the outward shock interacts with a shell of CSM material \citep{Chugai2004MNRAS.355..627C} that could be ejected by the progenitor during the pre-explosion activity.
The formation of this CDS would make the optical light curve fade faster, while the NIR one would remain bright, or even show an infrared excess, as observed in some fast declining SNe Ibn \citep{Pastorello2008MNRAS.389..131P, Mattila2008MNRAS.389..141M}.
Such a scenario was proposed also for the Type IIn SN 1998S \citep{Pozzo2004MNRAS.352..457P}.
However, we lack NIR observations to confirm this hypothesis.
It is also worth noting that the influence of the CSM on the light curve depends by many factors such as its mass, its density structure, and its distance from the progenitor. Some combination of these parameters can produce a faster declining light curve instead of a slower one.  

In contrast, some SNe IIn remained luminous for years, being powered by continuous interaction with the CSM (such as SNe 2005ip, 2006jd, 2010jl, 2015da, 2017hcc, \citealt{Stritzinger2012ApJ...756..173S, Ofek2019PASP..131e4204O, Tartaglia2020A&A...635A..39T, Moran2023A&A...669A..51M}).
Two of those long-lasting events (SNe 2010jl and 2015da) are also part of our sample, and we did not see precursor events before them. It is possible that the massive CSM necessary to sustain the SN for many years was produced decades before explosion, therefore missing from our data.

We mention a couple of caveats concerning the possible connection between the occurrence of precursors and the decline rates of the SN light curves:
\begin{enumerate}
    \item SNe without an observed precursor may also have experienced such events, but remained undetected in our data. This is especially true for the older SNe and for those with few pre-explosion images.
    \item A connection between the occurrence of pre-SN outbursts and the decline rates of the SN light curves would  not necessarily imply that these two phenomena are causally connected.
\end{enumerate}

\subsection{Occurrence rate of pre-SN outbursts} \label{sect:rate}

\begin{figure}
\includegraphics[width=1.0\columnwidth]{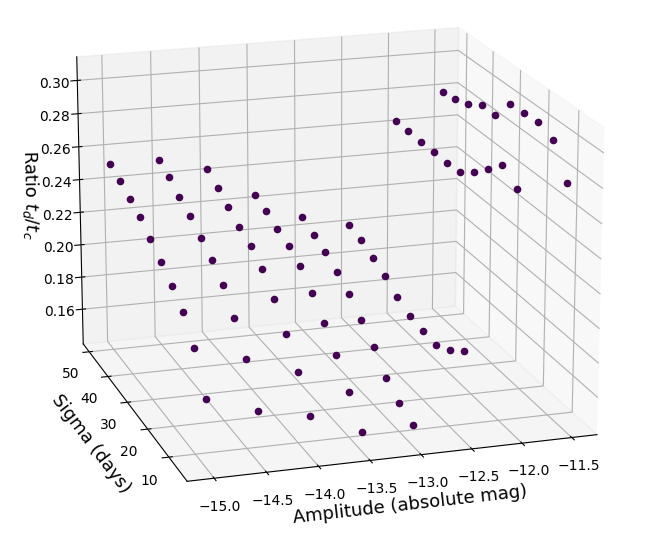}
\caption{3D plot of the $t_d$/$t_c$ ratio calculated for a grid of parameters of the Gaussian functions. The $a_p$ range from $-11.5$ to $-15$ $r$-band absolute magnitudes with steps of 0.5 mag, while $\sigma_p$ range from 5 to 50 days with steps of 5 days. The $t_d$/$t_c$ ratios have a median of 0.22, with a maximum of 0.30 and a minimum of 0.15.}
\label{fig:grid}
\end{figure}

\begin{figure}
\includegraphics[width=1.04\columnwidth]{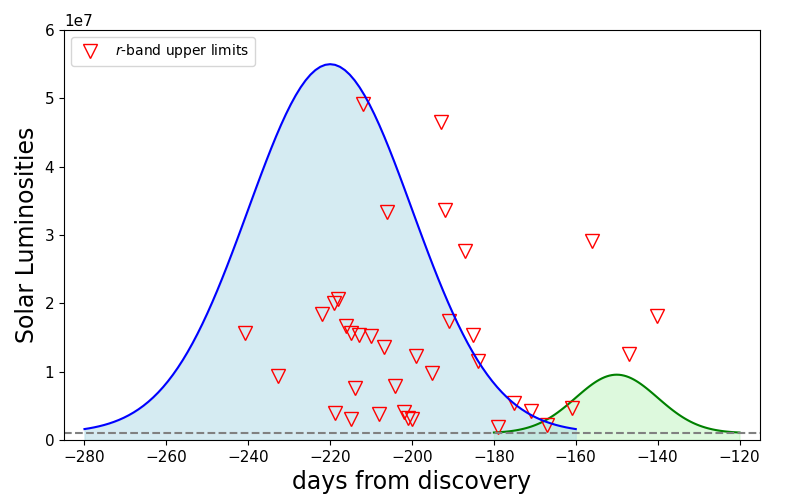}
\caption{Visualisation of the simulation used to estimate the rate of pre-SN precursors. The red triangles mark the upper limits (non-detections) of the pre-discovery images of a SN in the sample, converted into Solar luminosities.
The blue curve is a Gaussian with $a_p=-14.5$ $r$-band absolute magnitudes and $\sigma_p=20$ days, while the green curve is a smaller one with $a_p=-12.5$ mag and $\sigma_p=10$ days. The quiescent luminosity of the progenitor is set at 1$\times10^6$ \lsun, and is shown by the grey dashed line.}
\label{fig:gauss}
\end{figure}

In contrast with previous works (\citetalias{Ofek2014ApJ...789..104O}, \citetalias{Bilinski2015MNRAS.450..246B}, \citetalias{Strot2021ApJ...907...99S}), our research for pre-SN variability is not based on data regularly collected by a cadenced survey, such as PTF, KAIT/LOSS or ZTF, but rather on images from both the major surveys \textit{and} from observatories archives, whose temporal cadence is basically random for our purpose. The latter make impossible to obtain quantitatively statistically robust estimation of the intrinsic pre-SN outbursts occurrence rate and/or duration. Nevertheless, it is interesting to report the  first order estimation of the precursor events statistics that we can extract from the data.
As reported in Table \ref{tab1}, we detect precursor activity for 7 among the 27 SN~IIn belonging to our sample. Nevertheless, as pointed out in the footnotes of Table \ref{tab1}, two events (SN~2011A and PTF11qnf) are considered as SN Impostors \citep{DeJaeger2015ApJ...807...63D, Nyholm2017A&A...605A...6N}, while we suspect that SN 2018lkg is not a SN event (Appendix \ref{appendix}). Thus, if we remove those 3 objects from our list of \textit{bona-fide} SNe IIn, the size of our sample shrinks to 24, while the number of SNe with precursors is unchanged (7, for a fraction of 29\%).
For guiding our analysis on how the latter fraction compare with the intrinsic pre-SN outbursts occurrence rate, first of all, we need to obtain an estimation of the time during which we have images deep enough to detect a given precursor; we refer to this as our ‘control time ($t_c$)'. 
To estimate $t_c$ for each SN, we run a simulation\footnote{The \texttt{Python} code used to run the simulation with the Gaussians can be retrieved from Github: \url{https://github.com/Astronomo94/Gaussian-simulation}} that moves the maximum epoch of the Gaussian function through steps of one day across the time spanned between the oldest and the last observation prior to the SN explosion. Then, we count how many days at least one data-point in the $R$, $r$ or Clear filters is beneath the function, or - in other words - if the limiting flux of a certain observation is smaller than the expected flux of the function. As a consequence, if one upper limit is below the function with a certain maximum epoch, we would have spotted the outburst. The choice of the filters is dictated by the fact that most of the data are available in those filters. A visual representation of how the simulation works is shown in Fig. \ref{fig:gauss}.
With the same procedure, we also estimate another important parameter for our purpose, which is the duration of precursor activity ($t_d$). In this case the days are counted when at least one precursor detection is beneath the Gaussian function. With this approach, we do not distinguish if the points correspond to one long outburst or multiple short ones. Since both the estimates of $t_c$  and $t_d$ depend on the amplitude ($a_p$) and the width ($\sigma_p$) of the Gaussians, we constructed a grid of $a_p$ and $\sigma_p$ where the range of $a_p$ is from $-11.5$ to $-15$ mag (which are the faintest and brigthest $r$-band absolute magnitudes of the detected precursor events), with steps of 0.5 mag. The range of $\sigma_p$ is from 5 to 50 days (approximately the smallest and largest $\sigma_p$ of the Gaussian fits to our precursors), with steps of 5 days. It is worth noting that changing the assumed value of the background level has a negligible effect on those calculations, because it is one order of magnitude lower than the amplitudes of the Gaussian derived from fitting the precursors.  
Then, we sum the $t_d$ and $t_c$ times for all SNe in the sample for a specific $a_p$ and $\sigma_p$ and calculate the ratio $t_d$/$t_c$, which is plotted in Figure \ref{fig:grid}. The median ratio is 0.22, with a minimum and maximum values of 0.15 and 0.30, respectively. The overarching pattern indicates that the ratio smoothly increases with both the amplitude and the sigma of the Gaussian. Nevertheless, there is a significant increase of the ratio for $a_p$ lower than $-12.0$, which is due to the fact that, in this range of flux, the contribution to both $t_c$ and $t_d$ from the images retrieved by the DES survey start to be relevant. 
The values of $t_d$/$t_c$ give us a first indication that the fraction of SN~IIn for which we detect precursor activity in our sample is probably an underestimation of the real occurrence rate of pre-SN outbursts. Another insight on the same direction can be obtained considering that the average $t_c$\footnote{using an average Gaussian with $a_p=-13.5$ mag and $\sigma_p=15$ days} for the SNe with detected precursor is 474~days, while the same quantity for the SNe without a detected precursor is 179~days. With a similar $t_c$ for both groups, we would have a higher probability of detecting a precursor also in SNe for which no activity was observed from our dataset. We note that if we remove SN~2016ehw, whose $t_c$ is 1054~days, from the sub-sample of SNe without precursors, the previously mentioned average $t_c$ value shrinks further to 120~days. On the other hand, the case of SN~2016ehw can be taken as an indication that probably not all SNe~IIn display precursor before the explosion. In addition to the monitoring time, the absolute flux limit that the images can probe has a strong impact on the precursor statistics. In this regard, if we compute $t_d$/$t_c$ only for the images obtained by DES survey, the ratio increases to 0.76, and we detect outbursts in two of the three SNe monitored by them. Finally, considering the strong assumption we made to compute $t_d$ and $t_c$, i.e. that all the progenitor outbursts have the same evolution represented by a Gaussian function, we warn again the reader that all the above reported values have to be taken as order of magnitude estimations.

\section{Conclusions} \label{sec:conclus}
We searched for evidences of eruptive phases before the explosion of nearby SNe IIn. We constructed a sample of 27 SNe IIn situated within a DM of 34 mag, sufficiently close for  modern high-cadence surveys searching for new transients to be able to detect events with an absolute magnitude equal or brighter that $-14$ mag.
We looked at archival images of the SNe sites taken in the years prior to their discoveries, both from the surveys and the public archives of major Astronomical Observatories. Among the 27 objects, at least 7 show robust signatures of one or more eruptive episodes, indicating that the progenitors were experiencing a strong variability. 
As expected \citep{Smith2011MNRAS.415..773S}, the absolute magnitudes of these outbursts are between $-11$ to $-15$ mag, and the typical duration is a few weeks to months. 

We constructed the $g-r$ colour curves of the outbursts with multi-band observations.
The $g-r$ colours of the outbursts are red, in the range 0.5 to 1.0 mag. This is similar to that of LBVs during a giant eruption, when their atmospheres cool down and their spectra match those of F-type stars \citep{Humphreys1994PASP..106.1025H}. But it could also be an effect of a strong H$\alpha$ emission, a typical feature of SN impostors.
We verified with spectra of SN impostor events that this effect can only explain in part the observed red colours, hence the major contribution comes from the cooler photospheric temperature.

For testing whether the precursor activities had an effect on the post-explosion evolution, we measured the decline rates of the light curves of the SNe after maximum light from publicly available data. The cumulative distributions of the SNe~IIn with and without precursor activity show a marginally statistical difference, but contrary to expectations, the SNe that showed pre-explosion outbursts display much faster declines, indicating that the pre-explosion activity does not have a significant effect on the  SN~IIn light curve. The latter is in line with the results obtained by \citetalias{Ofek2014ApJ...789..104O} and \citetalias{Strot2021ApJ...907...99S}. 

Finally, we estimated the rate of the precursor events. We detect precursor activity in 29\% of bona-fide SN~IIn in our sample, which confirms that Type IIn is a class of SNe that frequently shows pre-explosion activity, in contrast with other SN types. Our estimation of the ratio between the time during which we detected activity with respect the time we could have detected provides an indication that the intrinsic occurrence rate of precursor events in the years prior to the terminal explosion is higher than that we obtain directly from the data, which may have been missed because of incomplete and too shallow data.

When the Legacy Survey in Space and Time at the Vera Rubin Observatory will start operations, it will scan the entire southern sky every 2 to 3 nights, reaching the 24th magnitude in a single 30 seconds exposure \citep{Ivezic2019ApJ...873..111I}. It will be able to detect much fainter outbursts from the progenitors of close-by SNe IIn, or will detect the brightest events at much larger distances. In both cases, it will enlarge by orders of magnitude the number of SNe with observed precursor activity.


\begin{acknowledgements}
\begin{small}
We thank the anonymous referee for their useful comments and suggestions, which improved the readability of the manuscript.
AR acknowledges financial support from the GRAWITA Large Program Grant (PI P. D’Avanzo).
AP and AR acknowledge the PRIN-INAF 2022 \textit{"Shedding light on the nature of gap transients: from the observations to the models"}. RD acknowledges funds by ANID grant FONDECYT Postdoctorado Nº 3220449. PTF was a scientific collaboration among the California Institute of Technology, Columbia University, Las Cumbres Observatory, the Lawrence Berkeley National Laboratory, the National Energy Research Scientific Computing Center, the University of Oxford, and the Weizmann Institute of Science.
ZTF is supported by the National Science Foundation and a collaboration including Caltech, IPAC, the Weizmann Institute for Science, the Oskar Klein Center at Stockholm University, the University of Maryland, Deutsches Elektronen-Synchrotron and Humboldt University, Lawrence Livermore National Laboratory, the TANGO Consortium of Taiwan, the University of Wisconsin at Milwaukee, Trinity College Dublin, and Institut national de physique nucléaire et de physique des particules. Operations are conducted by COO, IPAC and University of Washington.
Support for GP is provided by the Ministry of Economy, Development, and Tourism’s Millennium Science Initiative through grant IC120009, awarded to The Millennium Institute of Astrophysics (MAS).
This research has made use of the NASA/IPAC Extragalactic Database (NED) which is operated by the Jet Propulsion Laboratory, California Institute of Technology, under contract with the National Aeronautics and Space Administration.
We acknowledge ESA Gaia, DPAC and the Photometric Science Alerts Team (http://gsaweb.ast.cam.ac.uk/alerts).
\end{small}
\end{acknowledgements}

\bibliographystyle{aa} 
\bibliography{bib, non_ref} 

\begin{appendix} 
\section{SN (AT) 2018lkg} \label{appendix}

\begin{figure}[h]
\includegraphics[width=1.02\columnwidth]{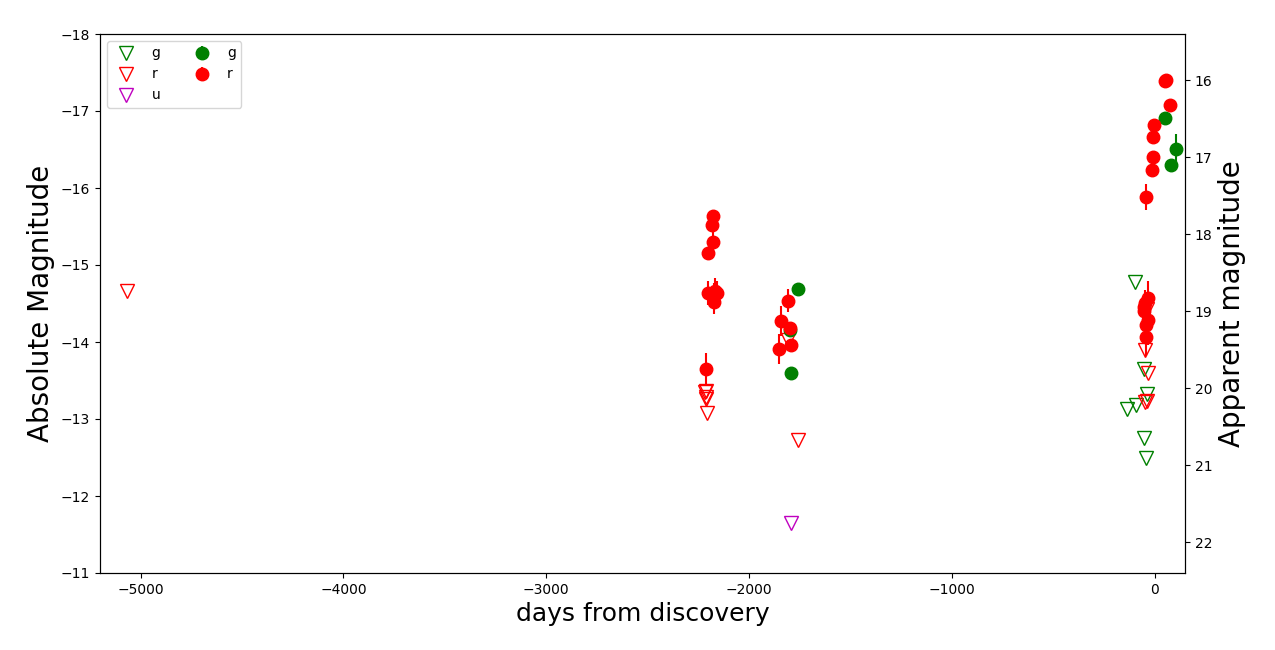}
\caption{
Pre-discovery $gr$ light curve of SN 2018lkg. In 2013 PTF detected a precursor event with absolute $M_r$ -16 mag, brighter than typical non-terminal events.}
\label{fig:18lkg}
\end{figure}

Analysing the sample, we note that SN 2018lkg is possibly an Active Galactic Nucleus, or a Tidal Disruption Event, rather than a Type IIn SN, because the transient sits in the centre of its host galaxy, and the multiple outbursts are unreasonably luminous to be produced by non-terminal stellar eruptions (being at nearly $M_r \sim -16$ mag, see Fig. \ref{fig:18lkg}). Furthermore, the classification spectrum \citep{Zhang2019ATel12358....1Z} reveals broad H features instead of the classical narrow lines of SNe IIn.

\begin{table*}
\caption{The Observatories locations and telescopes for which we consulted the public archives searching for pre-explosion images of SNe IIn. The observatories in the second part of the table do not provide additional useful images.}
\label{tab:observatories}
\begin{tabular}{lll}
\hline
Observatory/Archive & Location(s) & Telescopes \\
\hline
INT Group & La Palma & 4.2m WHT, 2.54m INT, 1.0m JKT \\
SMOKA Archive & Mauna Kea & 8.2m Subaru \\
GEMINI Archive & Mauna Kea, Cerro Pachon & 8.2m Gemini North and South \\
Canadian CADC & Mauna Kea & 3.6m CFHT \\
Esasky & Space & 2.4m HST \\
ESO Archive & La Silla, Paranal & 3.6m NTT \\
NOAO Archive & CTIO, Kitt Peak & 4.0m Victor Blanco, 4.1m SOAR and others \\
\hline
Asiago Archive & Mount Ekar & 1.82m Copernico, 0.67m Schmidt \\
TNG Archive & La Palma & 3.58m TNG \\
GTC Public Archive & La Palma & 10.4m GTC \\
LT Data Archive & La Palma & 2.0m LT \\
IA2 at INAF & Mount Graham & 2$\times$8.4m LBT \\
KOA & Mauna Kea & 2$\times$10m Keck \\
CAHA & Calar Alto & 3.5m and 2.2m Telescopes \\
\hline
\end{tabular}
\end{table*}

\end{appendix}

\end{document}